\documentclass[dvips]{article}
\usepackage{graphicx}
\usepackage{amssymb}
\usepackage{graphicx}
\usepackage{dcolumn}
\usepackage{amsmath}
\usepackage{lscape}
\usepackage{longtable}
\usepackage{subfigure}
\usepackage{color}

\begin{document}
\newcommand{\bd}{\begin{document}}
\newcommand{\ed}{\end{document}}
\newcommand{\bc}{\begin{center}}
\newcommand{\ec}{\end{center}}
\newcommand{\bfr}{\begin{flushright}}
\newcommand{\efr}{\end{flushright}}
\newcommand{\lt}{\left}
\newcommand{\rt}{\right}
\newcommand{\vs}{\vspace}
\newcommand{\hs}{\hspace}
\newcommand{\beq}{\begin{equation}}
\newcommand{\eeq}{\end{equation}}
\newcommand{\lb}{\linebreak}
\newcommand{\pb}{\pagebreak}
\newcommand{\mb}{\makebox}
\newcommand{\fb}{\framebox}
\newcommand{\mc}{\multicolumn}
\newcommand{\ben}{\begin{enumerate}}
\newcommand{\een}{\end{enumerate}}
\newcommand{\bit}{\begin{itemize}}
\newcommand{\eit}{\end{itemize}}
\newcommand{\ol}{\overline}
\newcommand{\un}{\underline}
\newcommand{\lefq}{\lefteqn}
\newcommand{\ba}{\begin{array}}
\newcommand{\ea}{\end{array}}
\newcommand{\beqa}{\begin{eqnarray}}
\newcommand{\eeqa}{\end{eqnarray}}
\newcommand{\beqas}{\begin{eqnarray*}}
\newcommand{\eeqas}{\end{eqnarray*}}
\newcommand{\bfg}{\begin{figure}}
\newcommand{\efg}{\end{figure}}
\newcommand{\bds}{\begin{displaymath}}
\newcommand{\eds}{\end{displaymath}}
\newcommand{\btb}{\begin{tabbing}}
\newcommand{\etb}{\end{tabbing}}
\bc {
\textbf{\huge $SU(1,1)$ solutions for the relativistic quantum particle in cosmic string spacetime} } \ec

\vs{1cm}

\bc
{\it \"Ozlem Ye\c{s}ilta\c{s}$^{*}${\footnote {e-mail : yesiltas@gazi.edu.tr}   \\
$^{*}$Department of Physics, Faculty of Science, Gazi University,
06500 Ankara, Turkey\\
\vspace{.16cm}

}} \ec \vs{1cm}
\begin{abstract}
We have studied a relativistic electron in the presence of a uniform magnetic field and scalar potential in the cosmic string spacetime. The  exact solutions of  the Dirac equation with a Coulomb-like scalar potential and linear vector potential through the gravitational fields  are found using $SU(1,1)$ Lie algebras.
\end{abstract}
\noindent {\bf keyword:}  relativistic particle, Dirac equation, curved spacetime, SU(1,1) \\


\section{Introduction}
The cosmic strings, which are known as one dimensional hypothetical topological defects \cite{cs1},  have been studied in a variety of physical systems. First predicted back in the 1970s \cite{cs1}, they are formed in symmetry breaking phase transitions in the early universe \cite{cs2}.
Kibble described the cosmic strings as forms of very concentrated mass-energy and they haven't yet been detected because of extremely high energies. This energy causes an extra gravitational force and topological defects are pioneers for the cosmic structure formations \cite{branden}.
Expansion during the first moments of the Universe leads to evolution of these strings which form an infinitely long network of strings and loops \cite{cs3}. At the same time, cosmic strings are also  analogous to those found in flux tubes in type-II superconductors \cite{cs4}, \cite{cs5}, \cite{cs6}, \cite{cs7}. When it comes to the geometry of a single cosmic string,  there are various studies done for the geometry generated by them \cite{cs8}, \cite{cs9}, \cite{cs10} and for the massless ones can be found in \cite{cs11}. The relativistic quantum dynamics of electric and magnetic dipoles in the presence of a topological defect is studied in \cite{cs12}, fermionic vacuum polarization in the cosmic string spacetime is analyzed in \cite{cs13}. The influence Aharonov-Casher effect on the Dirac oscillator in three different spacetimes including cosmic string spacetime was studied by Bakke and Furtado \cite{cs14}. The observation of the curvature of the spacetime at the atomic levels has been studying for the end of the last century. The solutions of the Dirac equation in the presence of gravitational fields in different spacetimes is being mentioned \cite{cs15}, \cite{cs16}, for example, in a quantum system with conical singularities \cite{cs17}, hydrogen atom in the gravitational fields generated by the cosmic string \cite{cs18}, exact solutions of Dirac equation in $(2+1)$ dimensional gravity \cite{cs19} and combination with position dependent mass models \cite{them}, for the massless fermions \cite{me}. Moreover, the Dirac equation for the physical problems has also been analyzed  in the point view of Lie algebra \cite{20}, supersymmetric quantum mechanics \cite{21}, \cite{quesne}, \cite{22} and exact bound state solutions \cite{23}, \cite{24}.

In this article, we have examined the relativistic quantum behavior of the massive charged fermions in the cosmic string spacetime within the Lie algebraic approach. The magnetic field which is parallel to the string is uniform \cite{cs15}, the scalar potential is chosen as Coulomb-like. Analytical solutions of the Dirac system is found in the context of $SU(1,1)$ Lie algebra \cite{liea}. In this work, section II includes the Dirac equation in curved spacetime and the diagonalization of the Dirac Hamiltonian. In section III, the reader can find the $SU(1,1)$  algebraic solutions of the eigenvalue equation.

\section{Dirac equation}
The local coordinates are $x^{i}=(t, r, \phi, z)$. The line element for the cosmic string spacetime is given by:
\begin{equation}\label{1}
    ds^{2}=dt^{2}-dr^{2}-\varrho^{2}r^{2}d\phi^{2}-dz^{2}
\end{equation}
where $\varrho$ is related to the deficit angle $\varrho=1-4\mu$, $\mu$ is the linear mass density,  $r\geq0, (t,z)~ \epsilon~ (-\infty, \infty)$, the azimuthal angle is defined within the range $\phi ~\epsilon~ [0, 2\pi]$. Thus, the metric tensor reads $g_{\mu \nu}=diag(1, -1, -\varrho^{2}r^{2}, -1 )$. A conical singularity appears in the geometry characterized by (\ref{1}) \cite{cs14}. The uniform magnetic field which is parallel to the string is shown as $\vec{B}=\vec{\nabla}\times\vec{A}=B_{0}\vec{k}$ and the Coulomb gauge is $\vec{A}=[0,A_{\phi}(r),0]$. $M$ is the mass of the particle, $\Gamma_{\mu}$ is the spinor affine connection, $S(r)$ is the scalar potential, and the Dirac equation is
\begin{equation}\label{2}
   [ i \gamma^{\mu}(x)(\nabla_{\mu}+i e A_{\mu})-(M+S(r))]\Psi(x)=0.
\end{equation}
And $\gamma^{\mu}(x)$ stands for the generalized Dirac matrices which satisfy $[\gamma^{\mu}(x), \gamma^{\nu}(x)]=2g^{\mu \nu}(x)$. The spinor connection can be written as
\begin{equation}\label{3}
    \Gamma_{\nu}=\frac{1}{8}e_{a\nu}\nabla_{\mu}e_{b}^{\nu}[\gamma^{a}, \gamma^{b}].
\end{equation}
 Constant Dirac matrices $\gamma_{a}$ and their relation with the non-flat ones is given as
\begin{equation}\label{4}
    \gamma^{\mu}(x)=e_{a}^{\mu}(x)\gamma_{a}
\end{equation}
where $e_{a}^{\mu}(x)$ are the tetrad fields satisfying $\eta^{ab}e_{a}^{\mu}(x)e_{b}^{\nu}(x)=g^{\mu \nu}$, also $\gamma_{0}=\left(
                                                                                                                              \begin{array}{cc}
                                                                                                                                \textbf{1} & 0 \\
                                                                                                                                0 & -\textbf{1} \\
                                                                                                                              \end{array}
                                                                                                                            \right)
$ and $\gamma_{a}=\left(
                    \begin{array}{cc}
                      0 & \sigma_{a} \\
                      -\sigma_{a} & 0 \\
                    \end{array}
                  \right)
$ where $\sigma^{a}$ are Pauli matrices, $\textbf{1}$ is the identity 2x2 matrix. The basis tetrad $e^{\mu}_{a}(x)$ and the inverse of it are chosen as \cite{cs15}
\begin{equation}\label{5}
   e^{a}_{\mu}(\textbf{x})=\left(
                  \begin{array}{cccc}
                    1 & 0 & 0 & 0 \\
                    0 & \cos \phi & \varrho r \sin \phi & 0 \\
                    0 & -\sin \phi & \varrho r \cos \phi & 0 \\
                    0 & 0 & 0 & 1 \\
                  \end{array}
                \right) ,~~~ e^{\mu}_{a}(\textbf{x})=\left(
                  \begin{array}{cccc}
                    1 & 0 & 0 & 0 \\
                    0 & \cos \phi & -\frac{\sin \phi}{\varrho r} & 0 \\
                    0 & \sin \phi & \frac{\cos \phi}{\varrho r} & 0 \\
                    0 & 0 & 0 & 1 \\
                  \end{array}
                \right).
\end{equation}
Hence, using (\ref{4}) and (\ref{5}), one obtains \cite{cs15}:
\begin{equation}\label{6}
    \gamma^{0}=\gamma_{0},~\gamma^{3}=\gamma_{3}, ~ \gamma^{i}=\left(
                                         \begin{array}{cc}
                                           0 & \sigma^{i} \\
                                           -\sigma^{i} & 0 \\
                                         \end{array}
                                       \right), ~~\sigma^1=\left(
                                                             \begin{array}{cc}
                                                               0 & e^{-i\phi} \\
                                                               e^{i\phi} & 0 \\
                                                             \end{array}
                                                           \right),~~ \sigma^2=-\frac{i}{\varrho r}\left(
                                                                                                    \begin{array}{cc}
                                                                                                      0 & e^{-i\phi} \\
                                                                                                      -e^{i\phi} & 0 \\
                                                                                                    \end{array}
                                                                                                  \right),~~~\sigma^3=\sigma_3.
\end{equation}
Using (\ref{3}), the nonzero component of the spin connection is obtained as $\Gamma_{\phi}=\frac{i(1-\varrho)}{2}\Sigma^{3}$ and then, the  Hamiltonian (\ref{2}) turns into
\begin{equation}\label{7}
    \mathrm{H}= -i\gamma_{0}\left(\gamma^{r}\frac{\partial}{\partial r}+\gamma^{\phi}\left(\frac{\partial}{\partial \phi}+\frac{i(1-\varrho)}{2\varrho r}\Sigma^{3}+i e A_{\phi}(r)\right)+\gamma_{3}\frac{\partial}{\partial z}+i(M+S(r))\right),
\end{equation}
where $\Sigma^{3}=\left(
                    \begin{array}{cc}
                      \sigma^{3} & 0 \\
                      0 & \sigma^{3} \\
                    \end{array}
                  \right)$.
The rotational symmetry leads to
            \begin{equation}\label{rot}
              [\mathrm{H}, J_3]=0
            \end{equation}
where  the total angular momentum operator in the $z$ direction  $J_3=L_{3}+S_3=-i\frac{\partial}{\partial_{\phi}}+\frac{1}{2}\Sigma^{3}$. Eigenvalue equations can be written for the operators as
\begin{equation}
  \mathrm{H}\Psi = E \Psi,~~  p_3 \Psi = k \Psi,~~  J_3 \Psi = j \Psi
\end{equation}
where $j=m+\frac{1}{2}, m=0, \pm 1, \pm 2,...$, k spans the whole space, $[\mathrm{H}, p_3]=0$, the momentum operator $p_3=-i\frac{\partial}{\partial z}$. Giving the wave function in the form of
\begin{equation}\label{8}
    \Psi(t,r,\phi,z)=\exp[-i E t+im\phi+ikz] \left(
                                               \begin{array}{c}
                                                 F_{+}(r) \\
                                                 -i F_{-}(r)e^{i\phi} \\
                                                 G_{+}(r) \\
                                                  G_{-}(r)e^{i\phi} \\
                                               \end{array}
                                             \right)
\end{equation}
and we obtain
    \begin{eqnarray}
     \frac{dG_{-}}{dr}+ (\frac{j}{\varrho r}+\frac{eA_{\phi}}{\varrho r} )G_{-}+kG_{+}+ (M+S(r)-E)F_{+} &=&0 \\
      \frac{dG_{+}}{dr}-(\frac{j}{\varrho r}+\frac{eA_{\phi}}{\varrho r} )G_{+}+kG_{-} + (M+S(r)-E)F_{-} &=& 0 \\
      \frac{dF_{-}}{dr}+(\frac{j}{\varrho r}+\frac{eA_{\phi}}{\varrho r} )F_{-}-kF_{+} + (M+S(r)+E)G_{+} &=& 0 \\
       \frac{dF_{+}}{dr}-(\frac{j}{\varrho r}+\frac{eA_{\phi}}{\varrho r} )F_{+}-kF_{-}+ (M+S(r)+E)G_{-} &=& 0.
     \end{eqnarray}
These four first order differential equations can be used to obtain
\begin{eqnarray}
 \frac{d^{2}F_{+}}{dr^{2}}+\left (-\frac{j(j-\varrho)}{\varrho^{2}r^{2}}+E^{2}-k^{2}-M^{2}\omega^{2}+\frac{2M\omega j}{\varrho r}-(M+S(r))^{2}\right)F_{+}+\frac{d S}{dr}G_{-} &=& 0 \\
  \frac{d^{2}G_{-}}{dr^{2}}+\left (-\frac{j(j+\varrho)}{\varrho^{2}r^{2}}+E^{2}-k^{2}-M^{2}\omega^{2}+\frac{2M\omega j}{\varrho r}-(M+S(r))^{2}\right)G_{-}+\frac{d S}{dr}F_{+} &=& 0
\end{eqnarray}
where $\omega=\frac{eA_{0}}{2M}$, $A_{\phi}(r)=-\frac{\varrho A_{0} r}{2}$ and $S(r)=\frac{s_1}{r}+s_2$. A simplified version of the pair of equations above can be expressed as \cite{cs15}
\begin{equation}\label{9}
    \Xi \left(
          \begin{array}{c}
            F \\
            G \\
          \end{array}
        \right)=0
\end{equation}
where $\Xi=B_1(r)\textbf{1}+B_2(r)\sigma^{3}+\frac{dS(r)}{dr}\sigma^{1}$ and we use $F_{+}=F$ and $G_{-}=G$. $B_1(r)$ and $B_2(r)$ are taken as
\begin{equation}\label{10}
    B_1(r)= \frac{d^{2}}{dr^{2}}-\frac{j^{2}}{\varrho^{2}r^{2}}+E^{2}-k^{2}+\frac{2M\omega j}{\varrho r}-(M+S(r))^{2},~~~~B_2(r)=\frac{j}{\varrho r^{2}}.
\end{equation}
The operator $\Xi$ can be diagonalized by $\mathrm{R}=\left(
                                               \begin{array}{cc}
                                                 \cos \frac{\theta}{2} & \sin \frac{\theta}{2} \\
                                                -  \sin \frac{\theta}{2} &  \cos \frac{\theta}{2} \\
                                               \end{array}
                                             \right)$ using $\bar{\Xi}=\mathrm{R} \Xi \mathrm{R}^{-1},~~\bar{\Xi} \left(
                                                                               \begin{array}{c}
                                                                                 \mathrm{R} F \\
                                                                                 \mathrm{R} G \\
                                                                               \end{array}
                                                                             \right)=0$ \cite{cs15}. Thus, we have,
 \begin{equation}\label{10}
    \bar{\Xi} \mathrm{R}=B_1(r) \left(
                                     \begin{array}{cc}
                                       1 & 0 \\
                                      0 & 1 \\
                                     \end{array}
                                   \right)+B_2(r) \left(
                                     \begin{array}{cc}
                                       \cos \theta & -\sin \theta \\
                                      - \sin \theta & -\cos \theta \\
                                     \end{array}
                                   \right)+\frac{d S}{dr}\left(
                                     \begin{array}{cc}
                                       \sin \theta & \cos \theta \\
                                       \cos \theta & -\sin \theta \\
                                     \end{array}
                                   \right).
 \end{equation}
 If we use
\begin{equation}\label{11}
  \tan \theta=- \frac{ s_1 \varrho}{j},~~~\tan \frac{\theta}{2}=-\frac{\varrho s_1}{\varrho \gamma+j},
\end{equation}
where $\gamma=\frac{1}{\varrho}\sqrt{j^{2}+\varrho^{2}s^{2}_1}$, then this operator becomes diagonal. So,  this procedure  allows us to give a couple of  second order differential equations as
\begin{equation}\label{12}
   \left(\frac{d^{2}}{dr^{2}}-\frac{j(j-\varrho)}{\varrho^{2} r^{2}}+\frac{2M(j\omega-\varrho s_1-\frac{\varrho s_1 s_2}{M})}{\varrho r}-s^{2}_1 (1-\frac{\varrho}{\varrho \gamma+j})\frac{1}{r^{2}}+E^{2}-k^{2}-M^{2}-2Ms_2-s^{2}_2\right)F(r)=0
\end{equation}
\begin{equation}\label{13}
    \left(\frac{d^{2}}{dr^{2}}-\frac{j(j+\varrho)}{\varrho^{2} r^{2}}+\frac{2M(j\omega-\varrho s_1-\frac{\varrho s_1 s_2}{M})}{\varrho r}-s^{2}_1 (1+\frac{\varrho}{\varrho \gamma+j} )\frac{1}{r^{2}}+E^{2}-k^{2}-M^{2}-2Ms_2-s^{2}_2\right)G(r)=0.
\end{equation}
\section{Factorization and $SU(1,1)$ algebra}
In order to show that the extended radial system possesses a symmetry \cite{gerry}, \cite{liea}, \cite{yes}; we shall factorize the left hand-side of (\ref{12}) and (\ref{13}). Then, if we use
\begin{eqnarray}
  \frac{j(j-\varrho)}{\varrho^{2}}+s^{2}_1(1-\frac{\varrho}{\varrho \gamma+j}) &=&\textbf{d}(\textbf{d}+\mathrm{a})  \\
  \frac{j(j+\varrho)}{\varrho^{2}}+s^{2}_1(1+\frac{\varrho}{\varrho \gamma+j}) &=& \textbf{d}(\textbf{d}-\mathrm{a})
\end{eqnarray}
and $\eta^{2}=-E^{2}+k^{2}+(M+s_2)^{2}$, $B=\frac{M(j\omega-\varrho s_1 -\frac{\varrho s_1 s_2}{M})}{\varrho }$ in (\ref{12}) and (\ref{13}), we have
\begin{eqnarray}\label{155}
  (-r^{2}\frac{d^{2}}{dr^{2}}+\eta^{2}r^{2}-2Br) F(r)&=& -\textbf{d}(\textbf{d}+\mathrm{a})F(r) \\ \label{14}
  (-r^{2}\frac{d^{2}}{dr^{2}}+\eta^{2}r^{2}-2Br) G(r) &=& -\textbf{d}(\textbf{d}-\mathrm{a}) G(r)\label{15}
\end{eqnarray}
where
\begin{eqnarray}
 \textbf{ d} &=&\sqrt{ s^{2}_1+\frac{j^{2}}{\varrho^{2}} }\\
  \mathrm{a} &=& -\frac{j+\frac{s^{2}_1\varrho^{2}}{j+\varrho \gamma}}{\varrho\sqrt{\frac{j^{2}}{\varrho^{2}}+s^{2}_1}}.
\end{eqnarray}
Our aim is to factorize (\ref{12}) and (\ref{13}). Let us give
\begin{equation}\label{16}
    (r\frac{d}{dr}+a_1 r+a_2)(-r\frac{d}{dr}+b_1 r+b_2)~\mathrm{y} =\lambda~ \mathrm{y}
\end{equation}
where we can define the constants if we compare (\ref{15}) and (\ref{16}). They are
\begin{equation}\label{17}
    a_1=b_1=\pm \eta, ~~~~a_2= -1 \mp \frac{B}{\eta},~~~~\lambda=a_2 (a_2+1)-\textbf{d}(\textbf{d}-\mathrm{a}).
\end{equation}
(\ref{15}) and (\ref{15}) can be also expressed as
\begin{equation}\label{18}
    (L_{\mp}\mp 1)L_{\pm} ~G_{\textbf{d}}=\left(\left(\frac{B}{\eta}\pm\frac{1}{2}\right)^{2}-\left(\textbf{d}-
    \frac{\mathrm{a}}{2}\right)^{2}+\frac{\mathrm{a}^{2}-1}{4}\right) G_{\textbf{d}}
\end{equation}
where $L_{\pm}=\mp r\frac{d}{dr}+\eta r-\frac{B}{\eta}$ and $\left(
                                                             \begin{array}{c}
                                                               F(r) \\
                                                               G(r) \\
                                                             \end{array}
                                                           \right)=\left(
                                                                     \begin{array}{c}
                                                                       \mathrm{y_{\textbf{d}+a}} \\
                                                                       \mathrm{y_{\textbf{d}}} \\
                                                                     \end{array}
                                                                   \right)
$. Let us re-write (\ref{15}) as
\begin{equation}\label{19}
    \Gamma_3 \mathrm{ y}_\textbf{d}=\frac{B}{\eta}\mathrm{ y}_\textbf{d},
\end{equation}
here $\Gamma_3=\frac{1}{2\eta}(-r\frac{d^{2}}{dr^{2}}+\eta^{2}r+\frac{\textbf{d}(\textbf{d}-\mathrm{a})}{r})$. Defining a pair of operators $\Gamma_{\pm}$ according to (\ref{18}) as below
\begin{equation}\label{20}
    \Gamma_{\pm}=\mp r\frac{d}{dr}+\eta r-\Gamma_{3}.
\end{equation}
It is straightforward to show that the operators $\Gamma_{\pm}$ and $\Gamma_{3}$ form the $SU(1,1)$ Lie algebra \cite{gerry}
\begin{equation}\label{21}
    [\Gamma_{\pm}, \Gamma_3]=\mp \Gamma_{\pm},~~~~[\Gamma_{+}, \Gamma_{-}]=-2 \Gamma_{3}.
\end{equation}
And the Casimir operator is given by
\begin{equation}\label{22}
    \Gamma^{2}=-\Gamma_{+}\Gamma_{-}+\Gamma^{2}_{3}-\Gamma_3.
\end{equation}
Hence, the eigenvalue equation becomes
\begin{equation}\label{23}
     \Gamma^{2} \mathrm{ y}_{\textbf{d}}=\textbf{d}(\textbf{d}-\mathrm{a})\mathrm{ y}_{\textbf{d}}.
\end{equation}
Differential operators for the upper component can be obtained by the shift $\textbf{d} \rightarrow \textbf{d}+\mathrm{a}$.
\subsection{\emph{Representations}}
The unitary irreducible representations of the $SU(1,1)$ algebra are represented by the equations given below
\begin{eqnarray}\label{C}
  J^{2}\mid \mu \nu> &=& \mu(\mu+1)\mid \mu \nu> \\ \label{c}
  J_3\mid \mu \nu> &=& \nu\mid \mu \nu> \\\label{nu}
  J_{\pm}\mid \mu \nu> &=& \sqrt{(\nu\mp \mu)(\nu \pm \mu \pm 1)}\mid \mu \nu\pm 1>
\end{eqnarray}
$J^{2}$ is known as Casimir operator, $\nu=\mu+w+1, w=0,1,2,...$ and $\mu> -1$. Therefore, using (\ref{19}), (\ref{23}), (\ref{C}) and  (\ref{c}), we can give $\mu$ and $\nu$ in terms of $\textbf{d}$ ,
\begin{equation}\label{24}
    \mu_{\textbf{d}}=\frac{1}{2}(-1+\sqrt{1-4\mathrm{a}\textbf{d}+4\textbf{d}^{2}})
\end{equation}
and
\begin{equation}\label{25}
   \nu_{\textbf{d}}=\frac{1}{2}(1+\sqrt{1-4\mathrm{a}\textbf{d}+4\textbf{d}^{2}})+n_{\textbf{d}}=\frac{B}{\eta},~~ n_{\textbf{d}}=0,1,2,...
\end{equation}
Hence we get
\begin{equation}\label{26}
    E_{n_{\textbf{d}}}=\left(k^{2}+(M+s_1)^{2}-
    \frac{M^{2}(\frac{e A_0 j-2\varrho s_1 s_2}{2M}-\varrho s_1)^{2}}{\varrho^{2}[\frac{1}{2}(1+\sqrt{1-4\mathrm{a}\textbf{d}+4\textbf{d}^{2}})+n_\textbf{d}]^{2}}\right)^{\frac{1}{2}}.
\end{equation}
For $n=0$, the ground-state function can be found as
\begin{equation}\label{27}
   \mathrm{ y}_{0}=N_{0}r^{\frac{1}{2}(1+\sqrt{1-4\mathbf{a}\textbf{d}+4\textbf{d}^{2}})} e^{-\frac{2Br}{1+\sqrt{1-4\mathrm{a}\textbf{d}+4\textbf{d}^{2}}}}.
\end{equation}
For higher states we propose a solution which is
\begin{equation}\label{28}
  G(r)=r^{\frac{1}{2}(1+\sqrt{1-4\mathrm{a}\textbf{d}+4\textbf{d}^{2}})}e^{-\eta r}\mathrm{g(r)}.
\end{equation}
Substituting (\ref{28}) in (\ref{15}) gives us
\begin{equation}\label{29}
    z \mathrm{g(z)}''+(2\delta-z)\mathrm{g(z)}'+\delta\frac{(1-\eta)}{\eta}\mathrm{g(z)}=0
\end{equation}
where $z=2\eta r$  and $\delta=\mu_{\textbf{d}}+1=\frac{1}{2}(1+\sqrt{1-4\mathrm{a}\textbf{d}+4\textbf{d}^{2}})$. The solution of (\ref{29}) is given in terms of  the confluent hypergeometric function $_{1}F_{1}[-n, \delta, z ]$ \cite{book} and we have
\begin{equation}\label{30}
    G_{n}(r)= N_{2} r^{\frac{1}{2}(1+\sqrt{1-4\mathrm{a}\textbf{d}+4\textbf{d}^{2}})}e^{-\eta r} ~~_{1}F_{1}[-n, \delta, \frac{r}{2\eta} ].
\end{equation}








\section{Conclusions}
The relativistic energy states of the Coulomb-like problem are found for the Dirac equation in the cosmic string spacetime with the contribution of the conical geometry of the defect. Here, equation (\ref{26}) which agrees with the results of the Coulomb potential \cite{ss} is a generalization of the energy levels in the cosmic string spacetime. The solutions of the eigenvalue equation are obtained using the advantage of the operators spanning the $SU(1,1)$ algebra, also it is seen that (\ref{155}) can be obtained from  (\ref{15}) by $\textbf{d}\rightarrow \textbf{d}+\mathrm{a}$ instead of a usual shift $\textbf{d}\rightarrow \textbf{d}+1$ and eigen-functions for (\ref{13}) are given in terms of confluent hypergeometric functions which can also expressed by Laguerre polynomials \cite{ss}. The quadratic surfaces of both positive and negative energy (\ref{26}) are sketched for the positive curvature which means the deficit angle is given in the range $0< \varrho <1$. In figure $(a)$, energy levels are hyperboloid with respect to $k$ and magnetic field strength for $\varrho=0.2$, figure $(b)$ shows that energy surfaces become flat planar when $\varrho=0.9$, figure $(c)$ shows the change in energy with respect to numbers $m$ and $k$ for $A_0=15$ and finally, figure $(d)$ shows the energy surface is a conical  with respect to $\varrho$ and $A_0$ for the quantum numbers $m=2$ and $n=3$. The readers may be interested in positive curvature cases which are important for the topological defects in crystalline solids \cite{son} for the future works.


\newpage

\begin{thebibliography}{99}
\bibitem{cs1} T. W. B. Kibble   J. Phys. A:Math. Gen. 9 1387 1976.
\bibitem{cs2} A. Vilenkin, E. P. S. Shellard, Cosmic Strings and Other Topological Defects,  Cambridge Monographs on Mathematical Physics, Cambridge University Press (July 31, 2000), ISBN-10: 0521654769 and ISBN-13: 978-0521654760.
    \bibitem{branden} R. H. Brandenberger, Int. J. Mod. Phys. A 9(13) 2117 1994.
\bibitem{cs3} T. Damour and A. Vilenkin, Phys. Rev. D 71, 063510 2005.
\bibitem{cs4} M. B. Hindmarshis  T. W. B. Kibble, Rep. Prog. Phys. 58  411 1995.
\bibitem{cs5} D. A.  Rogozin and L. V.  Zadorozhna, Astr. Nach. 334(9) 1051 2013.
\bibitem{cs6} B. Hartmann and F. Michel, Phys. Rev. D 86 105026 2012.
\bibitem{cs7} K. Miyamotoa and K. Nakayama, J. Cosmology and Astroparticle Phys. 2013(7) 012 2013.
\bibitem{cs8} A. G. Smith, in The Formation and Evolution of Cosmic Strings, Proceedings of the Cambridge Workshop, Cambridge,
England, edited by G. W. Gibbons, S. W. Hawking, and T. Vachaspati (Cambridge University Press, Cambridge, England,
1990).
\bibitem{cs9} V. B. Bezerra and E. R. Bezerra de Mello, Class. Quantum Grav. 11 457 1994; E. R. Bezerra de Mello Class. Quantum Grav. 11 1415 1994.
\bibitem{cs10} S. H. Hendi, Adv. in High En. 2014 697914 2014.
\bibitem{cs11} Maarten van de Meent, Phys. Rev. D 87(2)  025020 2013.
\bibitem{cs12} K Bakke, J. R. Nascimento and C. Furtado, Phys. Rev. D 78 064012 2008.
\bibitem{cs13} E. R. Bezerra de Mello, V. B. Bezerra, A. A. Saharian and A. S. Tarloyan, Phys. Rev. D 78 105007 2008.
\bibitem{cs14} K. Bakke and C. Furtado, Ann. Phys. 336 489 2013.
\bibitem{cs15} E. R. F. Medeiros and E. R. B. de Mello, Eur. Phys. J. C, 72 2051 2012.
\bibitem{liea} M. Salazar-Ram\'{i}rez, D. Mart\'{i}nez, R. D. Mota and V. D. Granados, J. Phys. A 43 445203 2010.
\bibitem{cs16} J. Carvalho, C. Furtado and F. Moraes, Phys. Rev. A 84 032109 2011.
\bibitem{cs17} V B Bezerra, Class. Quantum Grav. 8 1939 1991.
\bibitem{cs18} G. de Marques and V. B. Bezerra, Phys. Rev. D 66 105011 2002.
\bibitem{cs19} Y. Sucu and N. \"{U}nal, J. Math. Phys.  48 052503 2007.
\bibitem{them} G. D. Marques, V. B.  Bezerra and S. H. Dong, Mod. Phys. Lett. A 28(31)  1350137 2013.
\bibitem{me} \"{O}. Ye\c{s}ilta\c{s}, Advances in High En. Phys. 2014  186425 2014.
\bibitem{20} R. P. Mart\'{i}nez-y-Romero, H. N. N\'{u}\~{n}ez-Y\'{e}pez, A. L. Salas-Brito, Phys. Lett. A 339 259 2005.
\bibitem{21} H. Fakhri, N. Abbasi, J. Math. Phys.  42(6) 2416   2001.
\bibitem{quesne} C. Quesne, Int. J. Mod. Phys. A  6(9)  1567 1991.
\bibitem{22}  A. D. Alhaidari, Phys. Rev. A 65(4)  042109 2002.
\bibitem{23} M. Hamzavi, S. M. Ikhdair,  B. J. Falaye, Ann. Phys. 341 153 2014.
\bibitem{24} P. Roy and R. Roychoudhury, Mod. Phys. Lett. A  10(26) 1969 1995.
\bibitem{gerry} C. C. Gerry, Phys. Rev. A, 35(5) 2146 1987.
\bibitem{yes} \"{O}. Ye\c{s}ilta\c{s}, Phys. Lett. A, 377 49 2012.
\bibitem{book} M. Abramowitz and I. A. Stegun, Handbook of Mathematical Functions: with Formulas, Graphs, and Mathematical Tables (Dover Books on Mathematics) Dover Publications 1965, ISBN-10: 0486612724, ISBN-13: 978-0486612720 .
\bibitem{ss}  F. Cooper, A. Khare, U. Sukhatme, Supersymmetry in Quantum Mechanics, World Scientific Pub Co Inc (July 2002) ISBN-10: 9810246129
ISBN-13: 978-9810246129, pp:40.
\bibitem{son} C. Furtado, F. Moraes, Phys. Lett. A 188 394 1994.








\end{thebibliography}
\end{document}